\documentclass[aps,pra,preprint,showpacs]{revtex4-1}

\usepackage{amsmath}
\usepackage{amssymb}
\usepackage{graphicx}
\DeclareMathOperator\erf{erf}
\DeclareMathOperator\erfc{erfc}

\begin{document}

\title{Quantum capacitance in monolayers of silicene and
related buckled materials}
\author{S. Nawaz$^{\dag }$ and M. Tahir$^{\ast }$}
\affiliation{$^{\dag }$CNR-IOM Laboratorio TASC, Area Science Park, Basovizza, 34149 Trieste, Italy and International Centre for Theoretical Physics (ICTP), I-34014 Trieste, Italy, and CDL, Physics Division, PINSTECH, P. O. Nilore, Islamabad, Pakistan}
\affiliation{$^{\ast }$Department of Physics, Concordia University, Montreal, Quebec,
Canada H3G 1M8}

\begin{abstract}
Silicene and related buckled materials are distinct from both the
conventional two dimensional electron gas and the famous graphene due to
strong spin orbit coupling and the buckled structure. These materials have
potential to overcome limitations encountered for graphene, in particular
the zero band gap and weak spin orbit coupling. We present a theoretical
realization of quantum capacitance which has advantages 
over the scattering problems of traditional transport measurements. 
We derive and discuss quantum capacitance as a function of the Fermi 
energy and temperature taking into account electron-hole puddles through a Gaussian 
broadening distribution. Our predicted results are very exciting and 
pave the way for future spintronic and valleytronic devices.
\end{abstract}

\maketitle

\section{Introduction}

It is now well established that graphene is the ever first stable material
with two dimensional (2D) electronic structure. Graphene does not only carry the
exceptional electronic properties but also has enormous potential for
applications in various fields \cite{1}. However, its applications to
electronic industry are challenging due to the absence of an intrinsic band
gap and very weak spin-orbit coupling (SOC). Therefore, a race has begun in
search of other similar materials which could over rule the limiting factors
of graphene. Indeed, the first choice to explore was to study the close
relatives of carbon i.e., silicon, germanium and tin for the
stability of their 2D sheets \cite{2}. The motivation begins due to the
existence of an intrinsic energy gap in these materials and to their
larger ionic radii. Nevertheless the later lead to their 2D buckled
honeycomb lattice structures with enhanced SOC.

In recent years, there is a lot of experimental progress towards the
realization of stable silicene \cite{3,4}. Silicene is an excellent material
for electronic applications due to its compatibility with the existing
Si-based technology. Moreover, the realization of stable 2D structures of
silicene, germanene and stanene provides an exciting state of matter beyond
graphene. These layered materials open the possibility to explore
2D systems with strong SOC and structural buckling.
Additionally, the band gap of silicene is tunable by an external
perpendicular electric field applied along the buckling direction \cite{5,6}. In the
meanwhile, theoretical studies have predicted the stability of silicene on
different substrates such as graphene \cite{7,8}, boron nitride, silicon
carbide \cite{9} and solid argon \cite{10}. Recently, silicene and germanene
have been grown and realized to be stable at room temperature on gold and
silver surfaces as well \cite{11,12,13}. Nevertheless silicene field effect
transistors (FETs) have been demonstrated most recently \cite{14}. These materials are
expected to show exotic properties such as quantum spin- and valley-Hall
effects among many others \cite{15,16,17}. 
Although the focus of silicene research has been towards its transport properties, 
however, insight into its fundamental electronic properties and device physics still 
calls for knowledge about the capacitance-voltage (C-V) characteristics.

Quantum capacitance can be effectively used to probe the thermodynamical density of states (DOS).
These measurements not only provide an important way to
understand the fundamental electronic properties of materials but also have
potential utilization in device fabrication. For instance, the study of 
quantum capacitance in graphene has shown important implications on the design of 
FETs \cite{18,19,20,21,22,23}. Moreover, the
quantum capacitance has advantages over traditional transport measurements
whereas the latter are more
complicated and sensitive to scattering details. Furthermore, to improve 
the performance of FETs, the potential of silicene as a
channel material is creating much excitement due to strong SOC and 
tunable buckled structures. This is because of its excellent intrinsic transport 
features as well as the possibility of patterning device structures 
within top-down lithographical approach. The present work aims at 
determining the combined effects of SOC and perpendicular electric 
field on quantum capacitance of silicene and related 
buckled materials. Quantum capacitance is described by $C_{Q}=e^{2}D(E)$, 
where $D(E)$ represents the DOS, thus providing a quantitative description of
the DOS at the Fermi energy. Despite its capability of directly probing
electronic properties at finite temperatures, to the best of our knowledge,  
$C_{Q}$ has not been studied for silicene, germanene and stanene.

\section{Model Formulation}

We consider a monolayer silicene by an effective Hamiltonian in xy-plane. An
external perpendicular electric field is applied to the silicene sheet
taking into account the effects of SOC. Dirac fermions in buckled silicene obey
the 2D graphene-like Hamiltonian \cite{2,5,6}%
\begin{equation}
H_{s}^{\eta }=v(\eta \sigma _{x}p_{x}+\sigma _{y}p_{y})+\eta s\lambda \sigma
_{z}+\Delta \sigma _{z}  \label{1}
\end{equation}%
Here $\eta =+/-$ for $K/K^{\prime }$ valleys, $\Delta =2lE_{z}$, where $E_{z}$ is
the uniform electric field with $%
l=0.23$ \AA . In addition, ($\sigma _{x}$, $\sigma _{y}$, $\sigma _{z}$) is
the vector of Pauli matrices, $\lambda $ is SOC energy and $v$ denotes the
Fermi velocity of the Dirac fermions. The SOC energy for silicene, stanene and germanene is 3.9 meV,
29 meV and 43 meV, respectively \cite{2}. The up and down spin states 
are represented by $s=+1$ and $-1$, respectively. After
diagonalizing the Hamiltonian given in Eq.\ (1), we obtain the energy
eigenvalues%
\begin{equation}
E_{s}^{\eta }=\pm \sqrt{v^{2}\hslash ^{2}k^{2}+(\Delta +\eta s\lambda )^{2}}
\label{2}
\end{equation}%
We employ the delta function to represent DOS as%
\begin{equation}
D(E)=\sum_{\eta ,s}\delta (E-E_{s}^{\eta }),  \label{3}
\end{equation}%
The evaluation of DOS is straightforward and yields us%
\begin{equation}
D(E)=\sum_{\eta ,s}\frac{g_{s}g_{v}\left\vert E\right\vert }{2\pi (\hslash
v)^{2}}\theta \left[ y\right] .  \label{4}
\end{equation}%
Here $\theta \left[ y\right] \equiv \theta \left[ \left\vert E\right\vert
-\left\vert \Delta +\eta s\lambda \right\vert \right] $ is the Heaviside step
function, $g_{s}$ and $g_{v}$ are spin and valley degeneracy factors,
respectively. \ In the limits of zero SOC and electric field energies, the DOS
reduces to standard graphene expression \cite{19}.

Further, it has been demonstrated that graphene exhibits inhomogeneous landscapes of
electron-hole puddles around the Dirac point \cite%
{24,25,26,27,28}. Puddles are more probable to occur in gapless graphene 
as compared to a fully gapped silicene. However, the possible
existence of puddles and their effects in silicene layers are unexplored yet
and needs to be addressed.

We start by assuming that charged impurities, located in the substrate or
near the silicene surface, create a local electrostatic potential $\ V$
which fluctuates randomly about its average value.
The potential fluctuations are described by a
statistical distribution function $P(V)$ where $V=V(r)$ is the fluctuating
potential energy at the point $r\equiv (x,y)$ in the 2D silicene plane.
The probability $P(V)dV$ of finding the local electronic
potential energy within a range $dV$ about $V$ to a Gaussian form is 
approximated as \cite{29,33,34}

\begin{equation}
P(V)=\frac{1}{\sqrt{2\pi }\Gamma }\exp \left[ -\frac{V^{2}}{2\Gamma ^{2}}%
\right]  \label{5}
\end{equation}

Here $\Gamma $ is the standard deviation (or equivalently, the strength of
the potential fluctuation), which is used as an adjustable parameter to tune
the Gaussian broadening \cite{29}.

The potential fluctuations given by Eq. (5) affect the overall electronic
DOS. In our model we do not assume the size of the puddles to be
identical, instead we take the puddle size to be completely random, controlled
by the distribution function. We emphasize that our model provides an excellent quantitative 
approximation for the actual numerical calculations of puddle structures in
graphene \cite{30}. The characteristics of the puddles are determined by
both the sign and the magnitude of $V-E_F$, i.e., a negative (positive) $V-E_F$
indicates an electron (hole) region. A different approach utilizing equal
size puddles with a certain potential $V$ has been implemented to calculate
transport coefficients using a numerical transfer-matrix technique \cite%
{30,31,32}. In the presence of electron-hole puddles the $C_{Q}$ is
given by 

\begin{equation}
C_{Q}=\frac{e^{2}}{\sqrt{2\pi }\Gamma }\int_{-\infty }^{E}D(E)\exp \left[ -%
\frac{V^{2}}{2\Gamma ^{2}}\right] dV  \label{6}
\end{equation}%
Using Eq. (4) into Eq. (6) the $C_{Q}$ is simplified to%
\begin{equation}
C_{Q}=\frac{e^{2}D_{1}}{\sqrt{2\pi }\Gamma }\left[ \int_{-\infty }^{E}\sqrt{2%
}\Gamma E\exp (-x^{2})dx+\int_{-\infty }^{E}(\sqrt{2}\Gamma )^{2}x\exp
(-x^{2})dx\right] \theta \left[ y\right]   \label{7}
\end{equation}%
Here $D_{1}=\frac{g_{s}g_{v}}{2\pi (\hslash v)^{2}}$ and $x=\frac{V}{\sqrt{2}%
\Gamma }$. Using the definition of error function (
$\sqrt{\pi} [1+ \erf(x)]/2=\int_{-\infty }^{x}\exp [-(x)^{2}]dx$),
we arrive at the final result%
\begin{equation}
C_{Q}=e^{2}\sum_{\eta ,s}D_{1}\left[ \frac{E}{2}\erfc [-\frac{E}{\sqrt{2%
}\Gamma }]+\frac{\Gamma }{\sqrt{2\pi }}\exp -(\frac{E{}^{2}}{2\Gamma ^{2}})%
\right] \theta \left[ y\right]   \label{8}
\end{equation}%

\begin{figure}[b]
\includegraphics[width=0.45\columnwidth,height=0.3\columnwidth]{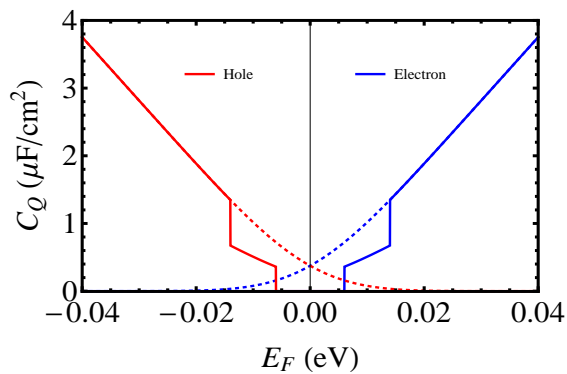}
\caption{Quantum capacitance as a function of the Fermi energy for
$\Gamma=10$ meV and $T=0 K$. The dotted and solid curves correspond to 
$\Delta=\lambda=0$ and $\Delta=10$ meV with $\lambda=4$ meV, respectively.}
\end{figure}

where $\erfc(x)$ is the complementary error function. Due to the electron-hole symmetry in the
problem, we only provide the formalism and equations for electron like
carriers and the hole part can be obtained simply by changing $E$ to $-E$. For the
case $\Gamma =0$, the system becomes homogeneous and $C_{Q}=e^{2}D_{1}E\theta \left[ y\right]$,
which is similar to that obtained in Eq. (4). In this case there is no carrier
density at the Dirac point ($E=0$) at zero temperature. Further, it is apparent that
in the presence of potential fluctuations, the $C_{Q}$ starts at finite
value at $E=0$ and approaches $e^{2}D_{1}E\theta \left[ y\right]$ in the limits of high-energy
and finite $\Gamma $. For the high-energy limits, the carriers are essentially
free. The electron density at finite temperatures increases due to thermal excitations
from valence band to conduction band, which is one of the important sources
of temperature-dependent transport at low carrier densities.

Eq. (8) shows dependence of $C_{Q}$ upon Fermi energy along with SOC and electric field 
energies. However, by taking into account the effects of low energy excitations 
like charge impurities or puddles, $C_{Q}$ is plotted in Fig. 1 as a function of
Fermi energy. It is evident from the results that the system
carries two energy gaps around the Dirac point (solid lines) due to the 
presence of both the applied electric filed and strong SOC energies. 
These gaps are tunbale by varying the perpendicular electric
field. One may notice that there is negligible puddle effect in silicene (solid line). 
In contrast, Eq. (8) reduces to gapless spectrum of graphene when the applied 
field and SOC energies are switched to zero (dottes lines) and hence we can see the puddle effect.

\begin{figure}[b]
\includegraphics[width=0.4\columnwidth,height=0.3\columnwidth]{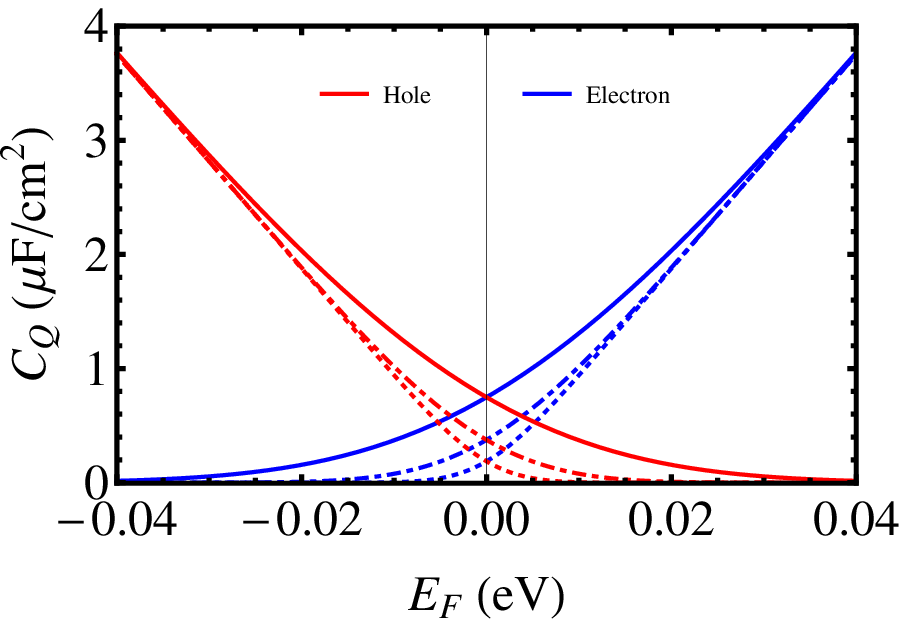}
\includegraphics[width=0.4\columnwidth,height=0.3\columnwidth]{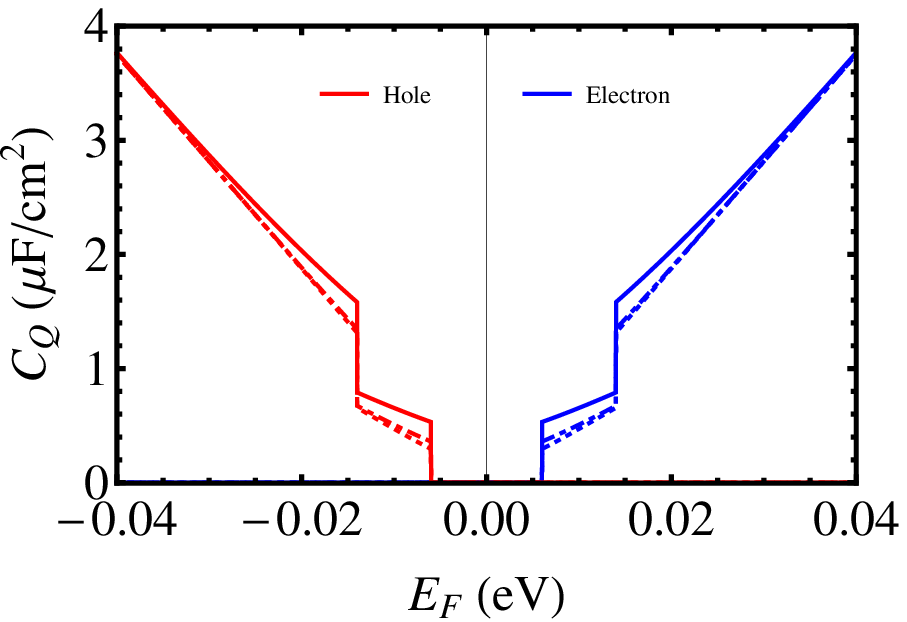}
\caption{Quantum capacitance as a function of the Fermi energy 
for $T=0 K$, $\Gamma=5$ meV (dotted curves), $\Gamma=10$ meV (dot-dashed curves), 
and $\Gamma=20$ meV (solid curves). 
The right panel is for $\Delta=\lambda=0$ and the left panel 
for $\Delta=10$ meV and $\lambda=4$ meV.}
\end{figure}

\begin{figure}[ht]
\includegraphics[width=0.6\columnwidth,height=0.45\columnwidth]{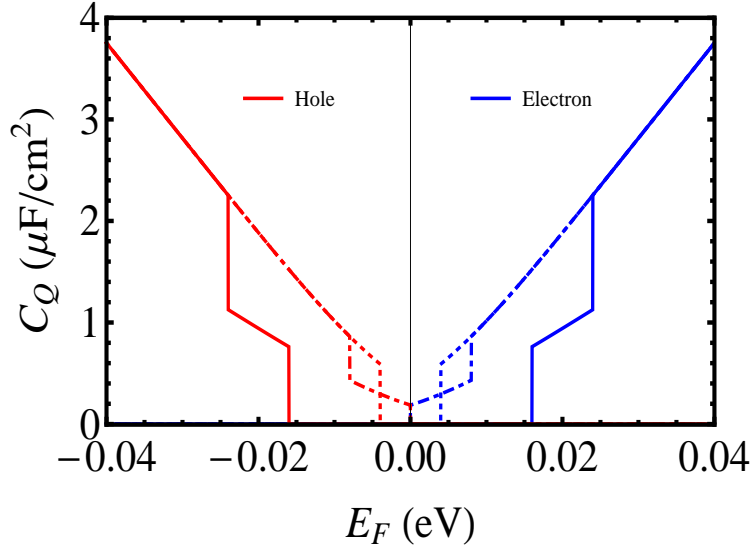}
\caption{Quantum capacitance as a function of the Fermi energy for $T=0 K$, 
$\Gamma=10$ meV and $\lambda=4$ meV. The dotted, dot-dashed, and solid curves correspond to 
$\Delta=0$ meV, $\Delta=4$ meV, and $\Delta=20$ meV, respectively.}
\end{figure}

To observe the expected puddle effect in silicene in detail, we show $C_{Q}$ 
as a function of the Fermi energy in Fig. 2. We find that the potential fluctuations 
caused by puddles in graphene affect the $C_{Q}$
considerably around the Dirac point (left panel). Contrary 
to this, the same effect seems significantly suppressed in silicene as shown in 
Fig. 2 (right panel). Further, it is interesting to note that the electric
field will lead to tuning of the $C_{Q}$ in the presence of puddles.

In silicene and related buckled materials there is an extensive interest towards the 
realization of quantum phase transitions \cite{5,6,15,16,17}. In these materials, the transitions 
are driven by tuning the perpendicular electric field in different regimes. 
In Fig. 3, we show quantum phase transitions for monolayer silicene
as a function of the Fermi energy. The dotted curve 
corresponds to topological insulating state, where $\lambda=4$ meV and $\Delta=0$ meV.
Here the width of the Gaussian broadening is fixed to $10$ meV.
The semi-metallic state for $\lambda=\Delta=4$ meV is shown in dot-dashed curve. The solid curve 
represents the band insulating state, where $\lambda=4$ meV and $\Delta=20$ meV.
We believe that these transitions would be easy to probe
through the measurements of DOS and the corresponding $C_{Q}$. This is because the $C_{Q}$ has advantages
over scattering complications in traditional transport measurements.

\begin{figure}[t]
\includegraphics[width=0.7\columnwidth,height=0.5\columnwidth]{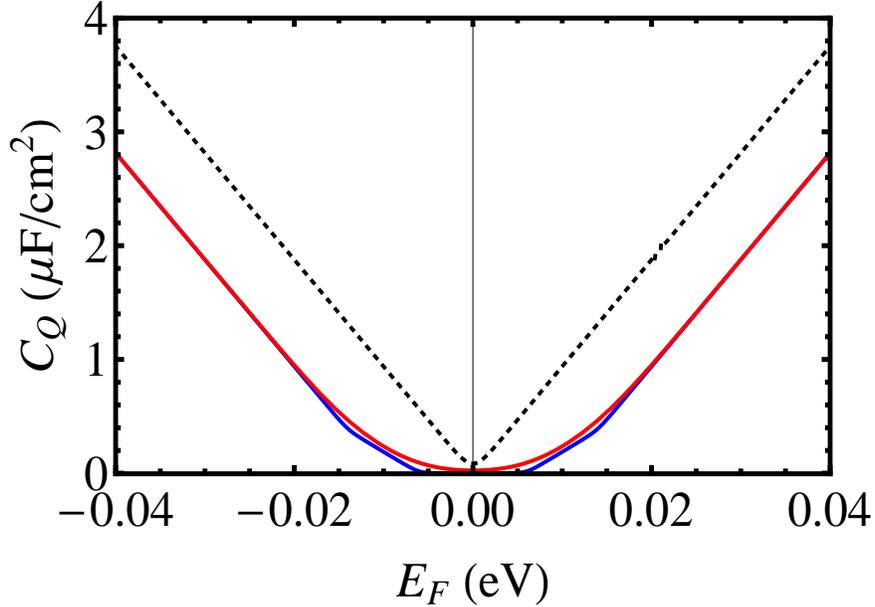}
\caption{Quantum capacitance as a function of the Fermi energy for
$\Gamma=1$ meV. The dotted and solid curves correspond to 
$\Delta=\lambda=0$ and $\Delta=10$ meV with $\lambda=4$ meV, respectively.
The dotted black and solid blue curves are for 5 K and solid red curve corresponds to 30 K.}
\end{figure}

Next we discuss the effects of temperature on $C_{Q}$ of the system under consideration. 
The $C_{Q}$ is defined as the derivative of the total net charge
of the system with respect to applied electrostatic potential \cite{19}. The total
charge is proportional to the weighted average of $C_{Q}$ at the Fermi level $%
E_{F}$. When the DOS as a function of energy is known, the $C_{Q}$ at finite temperature can be written as%
\begin{equation}
C_{Q}=e^{2}\int_{-\infty }^{+\infty }D(E)\left( -\frac{\partial f(E-E_{F})}{%
\partial E}\right) dE  \label{9}
\end{equation}%
One should note that the above expression is clearly valid for any temperature 
and Fermi energy. Here we show temperature dependent $C_{Q}$ in Fig. 4 as a function 
of the Fermi energy. We evaluate numerically $C_{Q}$ using Eq. (9) 
for a fixed value of Gaussian broadening $\Gamma=1$ meV. The blue and 
red solid curves correspond to $T=5 K$ and $30 K$, respectively. We note 
that by increasing the temperature, the effects of SOC and electric field are washed out.
In contrast, black dotted curve represents graphene in the limits of $\lambda=\Delta=0$ for 
a fixed temperature of $5 K$. To observe these transitions, the broadening of the 
levels must be less than the SOC and electric field energies. A similar value of 
the level broadening has been already achieved in high mobility graphene samples \cite{35}. 

The Hamiltonian in Eq. (1) can also be used to describe
germanene and stanene, which are honeycomb structures similar to silicene.
In these materials, the SOC energy is even stronger i.e. 43 meV (germanene) and 29 meV (stanene) \cite{2}.
Hence, the above analysis is fully applicable
to the systems like germanene and stanene as well. Our results imply that the level splitting can be
controlled by varying an external perpendicular electric field. We believe that the predicted
analysis is very exciting and opens the possibility for future spintronic and
valleytronic devices.

\section{Conclusion}

We have presented a theoretical model for the realization of quantum 
capacitance in monolayers of silicene and related buckled materials. For 
this, we employed the concept of broadened density of states taking into
account electron-hole puddles through a Gaussian distribution. 
The quantum capacitance has two jumps around the Dirac point which are 
representative of two energy gaps. These gaps show significant signatures of
quantum phase transitions as a competing consequence of SOC and 
electric field energies. Away from the Dirac point, the quantum 
capacitance becomes linear as a function of Fermi energy. As 
temperature increases, two jumps around the Dirac point 
are washed out, while quantum capacitance remains nearly unchanged otherwise.
In contrast to graphene, in the low temperature regime it seems that the
effects of the electron-hole puddles are suppressed. We believe
that these predictions open the possibility for tunable future spintronic and
valleytronic devices based on silicene and related buckled materials.
\\

{\bf Acknowledgments}: S. N. would like to acknowledge the financial support by the ICTP and CNR-IOM Trieste, Italy.\\ 

\noindent Electronic addresses: \\$^{\dag}$shadi665@gmail.com, $^{\star}$m.tahir06@alumni.imperial.ac.uk,\\

\end{document}